\documentclass[12pt,preprint]{aastex}
\shorttitle{Cool Gas in the Magellanic Stream}
\shortauthors{Matthews et al.}
\begin{document}
\title{COOL GAS IN THE MAGELLANIC STREAM}
\author{D. Matthews\altaffilmark{1}, L. Staveley-Smith\altaffilmark{2}, P. Dyson\altaffilmark{1}, and E. Muller\altaffilmark{3}} 
\altaffiltext{1}{Department of Physics, La Trobe University, Bundoora VIC 3086 Australia}
\altaffiltext{2}{School of Physics, University of Western Australia, Crawley, WA 6059, Australia}
\altaffiltext{3}{Physics and Astrophysics, Nagoya University ,Chikusa-ku, Nagoya 464-8602 Japan}
\begin{abstract} 
We present the first direct detection of cold atomic gas in the Magellanic Stream, through 21 cm line absorption toward a
background radio source, J0119 -- 6809, using the Australia Telescope Compact Array. Two absorption components were identified at
heliocentric velocities 218.6 km s$^{-1}$ and 227.0 km s$^{-1}$, with optical depths of $\tau\approx0.02$. The corresponding \ion{H}{1} emission region has a column density in excess of $2 \times 10^{20}$ cm$^{-2}$. The inferred spin temperature of the emitting gas is
$\sim70$ K. We failed to find cool gas in observations of three other radio continuum sources. Although we have definitively detected cool gas in the Stream, its spin temperature is higher than similar components in the LMC, SMC and Bridge, and its contribution to the total \ion{H}{1} density is probably lower. No corresponding $^{12}$CO(\textit{J} = 1 $\rightarrow$ 0) or dust appears to be associated with the cool gas, suggesting that the cloud is not forming stars.
\end{abstract}
\keywords{galaxies: Magellanic Clouds --- galaxies: interactions --- galaxies: formation --- radio lines: galaxies --- radio continuum: galaxies}
\section{Introduction}
The Magellanic System consists of the LMC, SMC, the Magellanic Bridge, which links the two, the Leading Arm, and the Magellanic Stream-- a ribbon of gas stretching behind the Magellanic Clouds for $\sim$100 degrees across the sky. Metallicity measurements \citep{gibson00,lu98} confirm the Stream is a low-metallicity environment, less than the Galaxy, M31 and M33, similar to the present day SMC (0.1 solar) and LMC (0.3 solar), but possibly higher than the Magellanic Bridge \citep{lehner08}. The formation and evolution of the Stream continues to be a contentious issue within the Magellanic community. The two dominant theories involve ram pressure stripping of LMC or SMC gas on their last passage through the ionized disk of the Milky Way \citep{mooredavis94,mastropietro05}, and the gravitational disruption of gas by the Galaxy in the last pericentric passage of the L/SMC \citep{connors06}. Strong support exists for both theories, although it is likely that both mechanisms play important roles \citep{hr94}. \citet{nidever07} put forward a more extreme suggestion that the Stream and Leading Arm are the results of supergiant shell blowouts in the LMC, with tidal and ram pressure forces shaping their morphologies. Controversially, recent proper motion measurements of the LMC and SMC by \citet{K1, K2} and follow-up simulations by \citet{Gurtina07} suggest that none of the existing models is viable, and propose instead that the System is either on its first passage about the Galaxy, or its orbital period and apogalactic distance are a factor of 2 larger than previously estimated. 

The presence of a cold atomic gas phase in \ion{H}{1} clouds is often accompanied by molecular cloud condensations, the precursor to star
formation. \ion{H}{1} 21 cm absorption and emission spectroscopy of the Clouds and Bridge have confirmed the existence of such cold atomic gas \citep{marxzimmer00, Dickey00, KobulDickey99}. The detection of CO adjacent to dense \ion{H}{1} cloudlets in the Bridge \citep{muller03} confirms that molecular cloud formation can occur in relatively tenuous, low-metallicity environments. However few studies have investigated the presence of cool gas in the Magellanic Stream \citep{Mebold91} . While the presence of such gas would not in itself indicate the presence of molecular clouds and star formation, it is still a significant step in determining the Stream's star-forming potential and the degree of fragmentation and collapse of its gas clouds.

The fine-structure lines of C and O (and, to a lesser extent, N, S and Si) provide the bulk of the cooling and improve the efficiency of star formation in the interstellar medium (ISM). Consequently, the ISM of metal poor-environments such as the Magellanic Stream will experience lower cooling rates, resulting in higher warm--cool gas ratios. \citet{marxzimmer00} showed that for the LMC the distribution and abundance of cool gas was determined by the local ISM conditions.

Given the apparent absence of dense gas and further dynamical perturbations, the ability for cool clouds to form in the Stream is markedly reduced, unless pressure confinement is significant \citep{stanim02}. The warm--cool gas phase mixture is not only dependent on the gas pressure and heavy element abundance, but also on the dust-to-gas ratio and the strength of the interstellar UV radiation field \citep{wolfire95}. A lower dust-to-gas ratio reduces shielding of molecules from photodissociation by the intense radiation field, allowing photodissociation fronts to penetrate deeper into clouds \citep{bot07}. \citet{wolfire95} showed that the range of pressures over which the two-phase medium is possible is narrower and the pressures in that range are higher. The higher pressure results in fewer cool phase regions in equilibrium and may explain the lower cool gas abundance in the SMC and Magellanic Stream. \citet{richter01} detected H$_2$ in the Stream toward the Seyfert galaxy Fairall 9 and suggested that for a Stream lifetime of $\sim 2$ Gyr, H$_2$ could form in situ in small compact cloudlets. \citet{braine01} suggest for typical \ion{H}{1} number densities only a few million years are required to transfer ~20$\%$ of atomic gas to molecular H$_2$. Alternatively, \citet{richter01} suggest that molecular gas could have survived intact during the host galaxy stripping process; however \citet{braine01} conclude this to be impossible, stating molecular gas in tidal streams is formed from condensing atomic gas and not pre-existing stripped clouds. 

The presence of cool atomic and molecular gas would have major implications for the fate of the Stream. Current estimates of
the mass of the Stream, $\sim 1.9 \times 10^8$ M$_{\odot}$ \citep{putman03}, only consider the neutral atomic hydrogen which is not abundant enough for gravitational collapse and the formation of a tidal dwarf galaxy \citep{braine01}. Recent numerical simulations \citep{recchi07, ducmirabel98} show that tails produced by tidal interaction are subject to fragmentation which may lead to star formation.

In this Letter we present the successful detection of cool atomic gas in the Magellanic Stream with the Australia Telescope Compact
Array (ATCA),\footnote{The Australia Telescope Compact Array is funded by the Commonwealth of Australia for operation as a National Facility managed by CSIRO.} and the result of a follow-up search for the $^{12}$CO(\textit{J} = 1 $\rightarrow$ 0) transition within the cool cloud with the Mopra telescope. In Section 2, we describe the emission and absorption observations and selection criteria for the four background continuum sources, and in section Section 3, we present the results of our investigations and discuss the implications of our results for theories of Magellanic Stream evolution.
\section{Observations}
The H75 configuration of the ATCA was used to observe a 250 deg$^2$ field of the Magellanic Stream centered on 1420 MHz. The observations were made in the period 2005 July 24-26, 27-30 and August 1-2. The field was sampled with 2618 separate pointings of the ATCA which were grouped into 17 regions, each containing 154 pointing centers. An integration time of 20 s per pointing was used, and each pointing center was visited approximately six times across a 10 hr observing day. The total integration time was therefore 120 s (2 minutes) per day, per pointing. A bandwidth of 4 MHz (corresponding to a velocity range of 840 km s$^{-1}$) was used, which was divided into 1024 channels per baseline for each of two polarizations. After Hanning smoothing the velocity resolution was 1.649 km s$^{-1}$. The rms sensitivity and angular resolution were 0.23 K and $7.4 \times 6.2$ arcmin$^2$ respectively.  The primary calibrator PKS B1934--638 was observed as a flux standard at the beginning of each observing day for $\sim$15 minutes and had an assumed flux density of 14.9 Jy at 1420 MHz. PKS B0252--712, used for the secondary (phase) calibrator (5.72 Jy at 1420 MHz), was visited for 5 minutes at the commencement of each observing session and revisited once per hour. Stray-radiation corrected Galactic All Sky Survey (GASS) data \citep{mcclure-griffiths09} were regridded to the same spatial and velocity dimensions as the ATCA data, and the two sets combined in the Fourier plane.

Following the identification of four radio sources in the background of relatively high column density Stream clouds (listed in Table
\ref{table1}), a follow-up search for the 21 cm line in absorption was performed with the ATCA. Two 12 hr observation runs
were made on 2007 February 7 and 8 using the 6A km array configuration. The correlator was configured to give a total bandwidth
of 4 MHz over 1024 channels, resulting in a velocity resolution of ~0.8 km s$^{-1}$. Only targets with column densities greater than 1.5
x10$^{20}$ cm$^{-2}$ directly in front of background continuum sources with flux greater than 300 mJy were selected (Figure
\ref{f1}). Two sources were observed daily, each for ~30 minutes at 70 minute intervals. At the beginning of both days the primary calibrator PKS B1934-638 was observed for 10 minutes, and before each source a secondary phase calibrator (PKS B0008-421 or PKS B0252-712) was tracked for 5 minutes. All data were reduced following standard procedures using the MIRIAD reduction package. The cleaned maps have a typical rms of 7.5 mJy beam$^{-1}$ per channel. Three objects (J0053-4914, J0110-6727, J0119-6809) are consistent with point sources at the resolution of our maps ($\sim 6$\arcsec). J0154-6800 is slightly resolved. Nevertheless, optical depth and spin temperature limits are still obtainable.

The UNSW-MOPS 3 mm spectrometer was used on the Mopra telescope to search for any molecular material that might be associated with the cool absorbing \ion{H}{1}. The UNSW-MOPS was configured in the narrow-band mode to perform dual-polarization position-switching observations of the $^{12}$CO(\textit{J\textit{}} = 1 $\rightarrow$ 0) transition at 115.271 GHz. A single pointing at the position of J0119-6809 was observed for 60 hr over the period 2007 October 1-2, 5-6, 8-10. The narrow-band mode configuration provides 138 MHz bandwidth over 4096 channels, resulting in a 360.45 km s$^{-1}$ velocity range with a resolution of 0.088 km s$^{-1}$. At the beginning of each session the calibrator was checked using the standard source M17 and showed that the system was stable to within 10$\%$ for the entire observing run. Every $\sim$ 75 minutes, the telescope pointing was recalibrated by observing nearby SiO maser sources RDor or UMen. Pointing corrections were repeated until errors less than 6$\arcsec$ were achieved. During each session system temperatures varied between approximately 420 and 800 K; however, only data obtained with Tsys less than 700 K were reduced. Data were bandpass-corrected, calibrated and smoothed over five channels using the ASAP package, resulting in a 3$\sigma$ brightness temperature of 36 mK and 0.44 km s$^{-1}$ velocity resolution.
\section{Spectral Analysis}
\begin{figure}
\epsscale{0.7}
\plotone{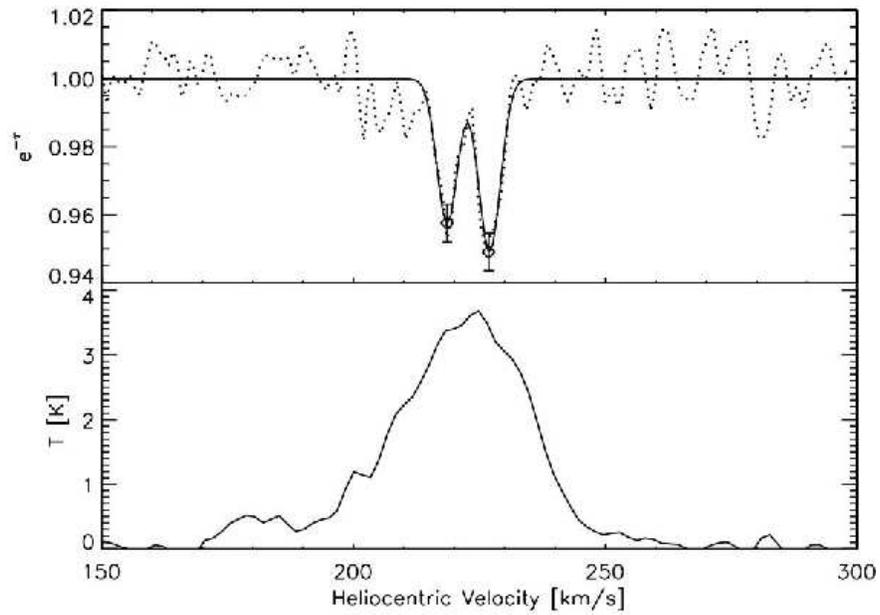}
\caption{Absorption (top) and emission (bottom) spectra toward the Magellanic Stream in the direction of the background continuum source
J0119-6809. The absorption spectrum was Hanning-smoothed to a channel width of 11.7 kHz (2.46 km s$^{-1}$). The 1$\sigma$ error in the fitted amplitude estimates are shown for both peaks.\label{f2}} 
\end{figure}

A 21cm absorption line was detected toward the continuum source J0119-6809. This is the first detection of cold atomic hydrogen in the Magellanic Stream. From the spectrum shown in Figure \ref{f2}, it is clear that the velocities of the absorption features at around 220 km s$^{-1}$ are in good agreement with the emission spectrum velocities which range from 200 to 250 km s$^{-1}$. This indicates the absorption arises from the same cloud. Two components are clearly identifiable in the absorption spectrum at 218.6 km s$^{-1}$ and 227.0 km s$^{-1}$ (Figure \ref{f2}). A sum of two Gaussians was fitted to the spectrum, and the fit components listed in Table \ref{table2}. A single Gaussian fit results in residuals that are too large compared with noise.

Upper limits on the \ion{H}{1} optical depths were obtained for the three sources without observable \ion{H}{1} absorption, and these are listed in Table \ref{table1}. Table \ref{table1} also lists the relevant source information. Although not representative of a physical temperature (since both the warm and cool components are present in any one channel), spin temperature ($T_{s}$) provides a means of estimating the temperature of the Stream's cold absorbing gas. \citet{Mebold91} use the emission brightness temperature and maximum optical depth of associated emission/absorption components to determine an upper limit for the cloud temperatures Table \ref{table1}. For the central velocity ($v_0$) of each component, the spin temperature is defined:
\begin{equation}
T_{s}(v_0)=\frac{T_B(v_0)}{(1-e{^{-\tau(v_0)}})}
\end{equation}
where $T_B$ is the brightness temperature of the emission, and $\tau$ is the optical depth of the absorption feature.  The calculated spin temperatures for components 1 and 2 are $79\pm 9$ K and $68\pm 6$ K, respectively  (Table \ref{table2}). These temperatures are low, and are an indication of the presence of at least some cool gas in the Stream. Nevertheless, the temperatures are higher than similar
measurements in the Magellanic Bridge (20 -- 40 K) by \citet{KobulDickey99}, the LMC (30 -- 40 K) by \citet{mebold97}, and the SMC (20 -- 50 K) by \citet{Dickey00}.  The column density of the surrounding warm \ion{H}{1} is similar to the column densities of the warm neutral envelope measured by \citet{KobulDickey99} (\textit{N}$_{H1}$ $\geq$ 10$^{20}$ cm$^{-2}$). Due to possible turbulent motions and warm gas components, the temperature deduced from linewidth alone is much higher than the spin temperature. 

Initial observations of molecular gas in the Bridge by \citet{smoker00} failed to detect $^{12}$CO(\textit{J} = 1 $\rightarrow$ 0)
down to an rms of 60 mK in this region. However, \citet{muller03} and \citet{mizuno06} both  detected CO within the Bridge in regions where where \textit{N}$_{HI}$ $\geq$ 10$^{21}$ cm$^{-2}$. Using emission maps of M81, \citet{taylor01} showed that CO regions can loosely trace tidally disrupted \ion{H}{1}. We therefore conducted a follow-up search for the $^{12}$CO(\textit{J} = 1 $\rightarrow$ 0) transition with the 22 m Mopra telescope in the direction of J0119-6809. No detection was apparent at the velocity of the \ion{H}{1} emission, with a 3$\sigma$ upper limit of 0.11 K.

IRAS far-infrared maps of the region also reveal no corresponding point sources above around 1 Jy. A search for extended emission using IRAS maps, from which the foreground Galactic dust emission had been subtracted using \ion{H}{1} data from GASS \citep{mcclure-griffiths09}, also reveal no excess emission above a 3$\sigma$ limit of around 0.24 MJy~sr$^{-1}$. Translated to a limit on the star formation rate (using the 0.1--100 M$_{\odot}$ formula of \citet{hop02}), the latter suggests an upper limit of $\sim 1.5 \times 10^{-5}$ M$_{\odot}$ yr$^{-1}$ over a region of size 220 pc (the GASS resolution at a distance of 50 kpc). No doubt these heuristic estimators of star formation break down at small spatial scales and at such small star formation rates, but it remains likely that star formation in this part of the Stream is exceedingly low and probably non-existent. \citet{braine01} estimate the mass of molecular gas required to form a TDG $\leq$ a few 10$^8$ M$_{\odot}$ leading us to conclude the Magellanic Stream will not condense, eventually forming a TDG; rather, clouds will continue to disperse and most likely fall onto the Galaxy. 
\section{Conclusion}
We made a search for evidence of star formation in the Magellanic Stream. For the first time in the Stream, we have located 21 cm \ion{H}{1} absorption. Of the four sources observed, we successfully detected cool atomic hydroden with a spin temperature of $\sim70$ K at the location J0119-6809. A follow-up search for molecular gas with the Mopra telescope failed to make a detection down to the resolution limit implying star formation could not be occurring in this region of the Stream. Only high \ion{H}{1} column density clouds ($2 \times 10^{20}$ cm$^{-2}$) were chosen for this investigation. Most Stream Clouds are much more tenuous than this, leading us to conclude that star formation is not presently an active process in the Magellanic Stream. 
\section{Acknowledgements}
\begin{figure*}
\centering
\epsscale{.75}
\plotone{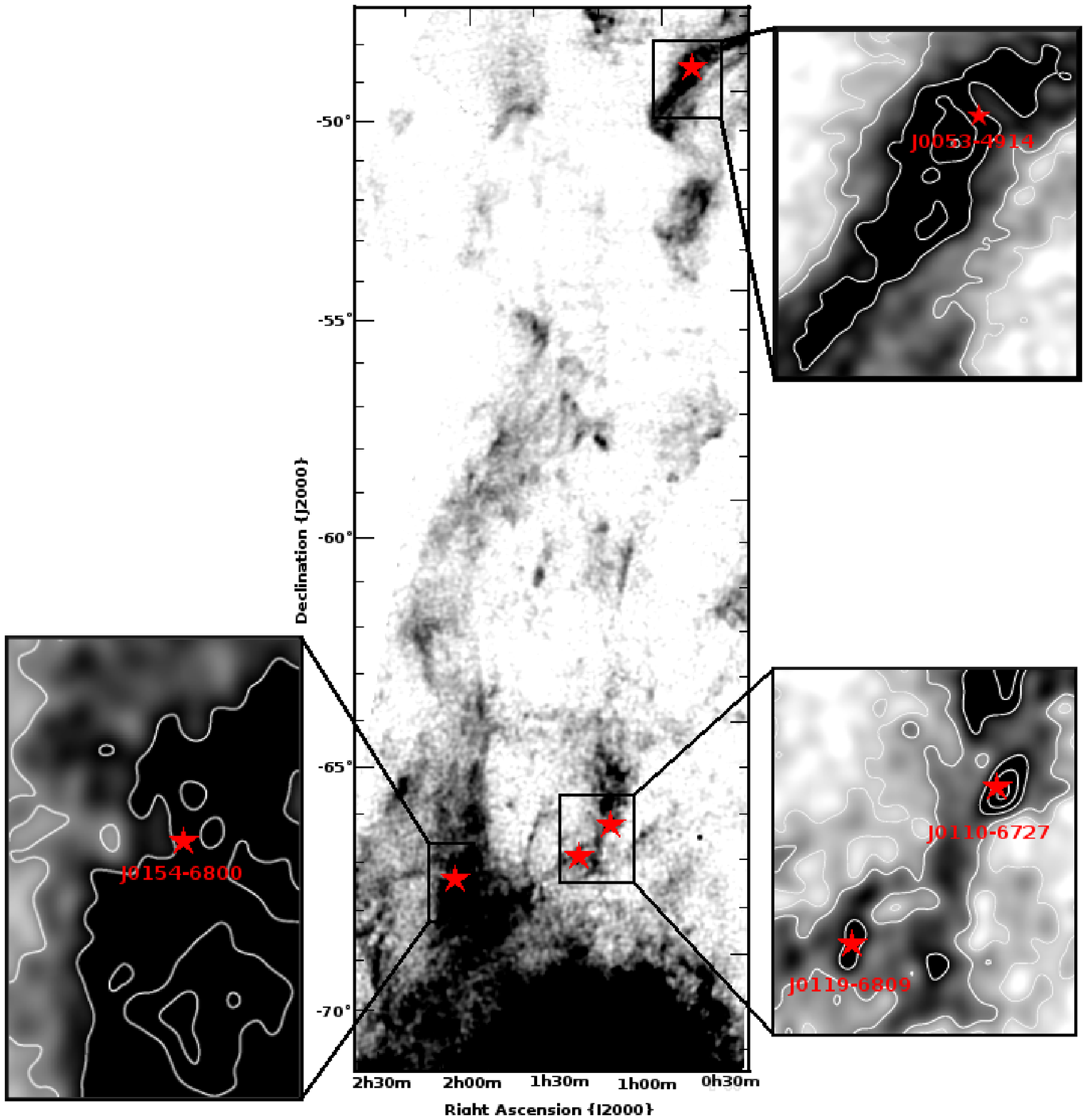}
\caption{\ion{H}{1} column density map of a region of the Magellanic Stream. Red stars define positions of the four candidate background continuum sources. Contours are at the 2, 3, 4, 5 and $6\times 10^{20}$ cm$^{-2}$ levels.\label{f1}}
\end{figure*}

We thank Naomi McClure-Griffiths for access to the preliminary GASS data used in Figure \ref{f1}. D.M. thanks Annie Hughes for very helpful discussions.
\begin{table*}
\begin{center}
\caption{\scriptsize{Observed background continuum source parameters}}
{\scriptsize
	\begin{tabular}{c c c c c c c c}
	\hline \hline
	Object & R.A. & Decl. & Flux density & N$_{HI}$   & 3$\sigma$ $\tau$ & $T_B$ & 3$\sigma$ $T_{s}$\\
 	& \tiny{(J2000)} & \tiny{(J2000)} &  \tiny{(mJy)} &  \tiny{($10^{20}$ cm$^{-2}$)} &  & \tiny{(K)} \\
	\hline
	J0053-4914 & 00 53 52.1 & $-49$ 14 03 & 743 & 2.2 & $<0.016$ & 5.30 & $>330$ \\
	J0110-6727 & 01 10 46.7 & $-67$ 27 56 & 358 & 3.7 & $<0.083$ & 7.46 & $>85$ \\
	J0119-6809 & 01 19 41.8 & $-68$ 09 34 & 659 & 2.3 & \multicolumn{3}{c}{--- see Table~\ref{table2} ---} \\
	J0154-6800 & 01 54 24.8 & $-68$ 00 41 & 355 & 2.4 & $<0.054$ & 5.02 & $>90$ \\
	\hline
	\end{tabular}
}
\label{table1}
\end{center}
\end{table*}
\begin{table}[!ht]
\vspace{-10pt}
\begin{center}
\caption{\scriptsize{J0119-6809 absorption-line parameters.}}
{\scriptsize
	\begin{tabular}{c c c c c c c}	
	\hline \hline
	Peak & $v_0$ & Height & FWHM & $T{_B}$  & $\tau$ & $T_s$($v_0$) \\
 	 & \tiny{(km s$^{-1}$)} & \tiny{(1-e$^{-\tau}$)} &  \tiny{(km s$^{-1}$)} &  \tiny{(K)} &   & \tiny{(K)} \\
	\hline
	1 & 218.6 & 0.043 $\pm$ 0.005 & 4.19 & 3.38 & 0.044 & 79 $\pm$ 9 \\
	2 & 227.0 & 0.051 $\pm$ 0.005 & 5.03 & 3.49 & 0.052 & 68 $\pm$ 6 \\
	\hline
\end{tabular}
}
\label{table2}
\end{center}
\vspace{-10pt}
\end{table}
\end{document}